\documentclass[letterpaper,titlepage,12pt]{article}
\usepackage{amssymb,amsmath,amsfonts,enumerate}

\def\clock{{\count0=\time
           \divide\count0 60
           \ifnum\count0<10 0\fi\the\count0
           \multiply\count0 -60 \advance\count0 \time
           :\ifnum\count0<10 0\fi \the\count0
         }}
\newcommand{\timestamp}{{\small\vbox{\hbox{\tt\jobname.tex}
\hbox{\the\day/\the\month/\the\year, \clock}}}}

\newcommand{\beq}{\begin{equation}}
\newcommand{\eeq}{\end{equation}}
\newcommand{\ben}{\begin{displaymath}}
\newcommand{\een}{\end{displaymath}}
\newcommand{\beqa}{\begin{eqnarray}}
\newcommand{\eeqa}{\end{eqnarray}}
\newcommand{\bea}{\begin{eqnarray}}
\newcommand{\eea}{\end{eqnarray}}
\newcommand{\bean}{\begin{eqnarray*}}
\newcommand{\eean}{\end{eqnarray*}}
\newcommand{\ba}{\begin{array}}
\newcommand{\ea}{\end{array}}
\newcommand{\bi}{\begin{itemize}}
\newcommand{\ei}{\end{itemize}}

\newcommand{\ie}{{\it i.e.,\,}}
\newcommand{\eg}{{\it e.g.,\,}}

\newcommand{\lp}{\left(}
\newcommand{\rp}{\right)}

\numberwithin{equation}{section}
\begin{document}

\begin{titlepage}
\begin{flushright}
%\timestamp
DCPT-12/05
\end{flushright}

\vskip 2.5cm
\begin{center}
{\bf\LARGE{Derivation of the blackfold effective theory %equations
}}
%\vskip 0.12cm
%{\bf\LARGE{from first principles}} 
\vskip 1.5cm
{\bf 
Joan Camps$^{a}$,
Roberto Emparan$^{b,c}$
}
\vskip 0.5cm
\medskip
\textit{$^{a}$ Centre for Particle Theory \& Department of Mathematical Sciences}\\
\textit{Science Laboratories, South Road, Durham DH1 3LE, United Kingdom}\\
\smallskip
\textit{$^{b}$Instituci\'o Catalana de Recerca i Estudis
Avan\c cats (ICREA)}\\
\textit{Passeig Llu\'{\i}s Companys 23, E-08010 Barcelona, Spain}\\
\smallskip
\textit{$^{c}$Departament de F{\'\i}sica Fonamental and}\\
\textit{Institut de
Ci\`encies del Cosmos, Universitat de
Barcelona, }\\
\textit{Mart\'{\i} i Franqu\`es 1, E-08028 Barcelona, Spain}\\

\vskip .2 in
\texttt{joan.camps@durham.ac.uk, emparan@ub.edu}

\end{center}

\vskip 0.3in

\baselineskip 16pt
\date{}

\begin{center} {\bf Abstract} \end{center} 

We study fluctuations and deformations of black branes over length
scales larger than the horizon radius. We prove that the Einstein
equations for the perturbed $p$-brane yield, as constraints, the
equations of the effective blackfold theory. We solve the Einstein
equations for the perturbed geometry and show that it remains regular on
and outside the black brane horizon. This study provides an \textit{ab
initio} derivation of the blackfold effective theory and gives explicit
expressions for the metrics near the new black holes and black branes that
result from it, to leading order in a derivative expansion.

\vskip 0.2cm

\noindent 
\end{titlepage} \vfill\eject

\setcounter{equation}{0}

\pagestyle{empty}
\small
%\tableofcontents
\normalsize
%\newpage
\pagestyle{plain}
\setcounter{page}{1}

\newpage

\section{Introduction}

The blackfold approach is a worldvolume effective theory for the
dynamics of black branes \cite{Emparan:2009cs,Emparan:2009at}. When the
worldvolume of the brane is spatially compact, it provides an efficient
tool to study new higher-dimensional black holes in regimes in which
their horizons possess two widely separated length scales
\cite{Emparan:2009vd}. It has also proved useful for understanding the
stability of black holes and black branes under long-wavelength
perturbations \cite{Emparan:2009at,Camps:2010br}.

This effective theory captures the dynamics of a black brane when it
deviates away from the flat uniform state over a scale $R$ much longer
than the brane thickness $r_0$. This can happen either because the brane
becomes inhomogenous, with thickness varying along the worldvolume
(intrinsic fluctuations) or because the worldvolume bends in the
background spacetime (extrinsic fluctuations). The dynamics of the brane
is conveniently expressed in terms of its effective stress-energy tensor
$T_{ab}$, which is computed in the region $r_0\ll r\ll R$ where the
gravitational field is weak. In \cite{Emparan:2009at}, using an
argument based on general covariance, the effective theory equations
were determined to be
\beq\label{inteqs}
D_a T^{ab}=0\,,
\eeq
for intrinsic fluctuations, where $D_a$ is the covariant derivative for
the worldvolume metric, and
\beq\label{exteqs}
K_{ab}{}^\rho T^{ab}=0\,,
\eeq
for extrinsic fluctuations, where $K_{ab}{}^\rho$ is the extrinsic
curvature tensor for the embedding of the brane worldvolume in the
background spacetime. The equations \eqref{inteqs} are of hydrodynamic
type, while \eqref{exteqs} are elasticity equations for a brane:
black branes behave as liquids under strains parallel to their
worldvolume, and like elastic solids under strains orthogonal to them.

These equations have been shown to describe correctly several properties
of black holes and black branes that were previously known, and have also
yielded new solutions and new insights. However, in order to more solidly ground
the theory it is necessary to

\begin{enumerate}

\item Derive the blackfold equations \eqref{inteqs} and \eqref{exteqs}
directly from the Einstein equations that describe a black $p$-brane with
generic perturbations of length scale $R\gg r_0$.

\item Prove that the horizon of the black $p$-brane remains regular under
these perturbations.

\end{enumerate}
The purpose of this paper is to establish these two points. 

The problem is a more familiar one in the case of $p=0$. It is well
known that, to the extent that a black hole can be seen as a point
particle, it must follow a geodesic worldline \cite{Poisson:2011nh}.
There are two sorts of arguments for this: (i) general covariance and
minimal coupling (\ie\ lowest derivative order) imply that the equations
of motion of the particle are uniquely fixed to be those of timelike
geodesics; (ii) the Einstein equations for a trajectory that is a small
perturbation of the static black hole solution yield the geodesic
equation as a constraint, and then the remaining Einstein equations
admit a unique solution that is regular on and outside the perturbed
horizon. We may regard (i) as abstract generic derivations, and (ii) as
`microscopic' or \textit{ab initio} derivations. 
The path to the blackfold equations in ref.~\cite{Emparan:2009at}
followed the abstract generic approach of (i). Our purpose here is to
derive these equations in the more detailed manner of (ii). Moreover,
just like geodesic motion describes the dynamics of any small particle,
not necessarily a black hole, similarly the extrinsic equations
\eqref{exteqs} are also applicable for all thin $p$-branes, not only
black ones\footnote{Eqs.~\eqref{exteqs} were first obtained from
covariance arguments of type (i) in \cite{Carter:2000wv}.}. Our proof of
these equations will indeed have this generality. The study of
regularity of the perturbed solution is instead theory-specific, in the
present case to neutral black $p$-branes.

In the next section we set up our study of generic long
wavelength perturbations, and argue that intrinsic and extrinsic
perturbations decouple at the leading order in a derivative expansion.
Intrinsic perturbations have been solved already in \cite{Camps:2010br}.
Then we proceed to set up the problem for extrinsic perturbations by
constructing Fermi normal coordinates adapted to $p$-branes. In section
\ref{sec:exteqs} we derive the extrinsic equations \eqref{exteqs} from
the gravitational dynamics of a generic $p$-brane. In section
\ref{sec:perthor} we solve the Einstein equations for a black $p$-brane
under extrinsic perturbations. In section \ref{sec:horreg} we show that
the horizon remains
regular. Section \ref{sec:completemetric} puts together in a covariant
manner the results for the metric of a black brane under a generic
long-wavelength perturbation. Section \ref{sec:discuss} contains a final
discussion of our results. As in previous works, for a $p$-brane in
$D$-dimensional spacetime we find convenient to introduce
\beq
n=D-p-3\,.
\eeq

\section{Set up}
\label{sec:setup}

\subsection{Long-wavelength perturbations of black branes}
We write the metric for a black $p$-brane in the form
\beq\label{blackp}
ds^2=\lp \eta_{ab}+\frac{r_0^n}{r^n}u_a u_b\rp d\sigma^a d\sigma^b+ 
\frac{dr^2}{1-\frac{r_0^n}{r^n}}
+r^2 d\Omega_{(n+1)}^2\,,
\eeq
where the coordinates $\sigma^a$ span the worldvolume of the $p$-brane
 with the Minkowski metric $\eta_{ab}$, $a,b=0,\dots p$. We have
introduced a constant worldvolume vector $u^a$ that characterizes the
local worldvolume velocity and satisfies $\eta_{ab}u^a u^b=-1$. 
In this
form, the parameters of the solution are the horizon radius $r_0$ and
the components of the velocity $u^a$. In addition to these, there are
$D-p-1$ coordinates for the position of the black brane in the transverse
space, $X^\perp$, which in \eqref{blackp} are fixed to the origin. They
can be made explicit by a shift of coordinates in the transverse
directions, but this results in cumbersome expressions that we will not
need. 

Constant shifts of these parameters still give solutions to the Einstein
equations. Our purpose is to describe solutions in which the parameters
vary slowly along the worldvolume. In a manner
patterned after \cite{Bhattacharyya:2008jc}, we write the modified geometry 
as
\beqa\label{nearpbrane}
ds^2
&=&
\left(\gamma_{ab}\left(X^\mu(\sigma)\right)+\frac{r_0^n(\sigma)}{r^n}u_a(\sigma)
u_b(\sigma)\right)d\sigma^a d\sigma^b+
\frac{dr^2}{1-\frac{r_0^n(\sigma)}{r^n}}+r^2
d\Omega^2_{(n+1)}\nonumber\\
&&+h_{\mu\nu}(x)dx^\mu dx^\nu\,,
\eeqa
where $\gamma_{ab}\left(X^\mu(\sigma)\right)=g_{\mu\nu}\partial_a
X^\mu\partial_b X^\nu$ is the metric induced on the worldvolume by the
embedding $X^\mu(\sigma)$ in the background metric
$g_{\mu\nu}$.\footnote{The transverse coordinate dependence is made more
explicit in section~\ref{subsec:fermi}.} The additional terms $h_{\mu\nu}$ are
of the same order as the derivatives of $\ln r_0$, $u^a$ and $X^\perp$,
and their presence is required in order to have a solution of Einstein's
equations. One of our aims is to compute these corrections and
prove that the horizon remains regular in the
corrected geometry. 

We shall work to first order in the derivative expansion. This means
that, locally, at any given point in the worldvolume the fluctuations
are small and we only keep terms linear in the perturbations. 
This local linearization of the perturbations is essential for our
analysis, as it allows us to decouple the intrinsic fluctuations (\ie of
$r_0$ and $u^a$, and the intrinsic curvature of the metric) from the
extrinsic fluctuations (of $X^\perp$), and therefore we
can discuss them separately.

\subsection{Intrinsic perturbations} The study of long-wavelength
fluctuations of $r_0$ and $u^a$, keeping the worldvolume metric flat,
has been done in \cite{Camps:2010br}. It has been shown that, to first
order in the derivative
expansion, the Einstein constraint equations $R^r{}_b=0$ require that
eqs.~\eqref{inteqs} be satisfied, where
\beq\label{Tab}
T^{ab}=\frac{\Omega_{(n+1)}}{16\pi G}r_0^n\left( n u^a u^b- \eta^{ab}\rp
\eeq
is the stress-energy tensor of the black brane measured in the asymptotically
flat region $r\to\infty$. This stress-energy tensor can be obtained
from an ADM-type
prescription \cite{Myers:1999psa}, or equivalently, from the Brown-York
quasilocal stress-energy tensor \cite{Brown:1992br}.

Since \eqref{Tab} can be regarded as the stress-energy tensor of a
perfect fluid, eqs.~\eqref{inteqs} are hydrodynamic Euler
equations. Ref.~\cite{Camps:2010br} then showed that, for any
solution of these $(p+1)$-dimensional equations, a solution of the
$D$-dimensional Einstein equations for $h_{\mu\nu}$ can be constructed
to leading order in derivatives, such that the horizon of the perturbed
black brane remains regular. Thus, the analysis in \cite{Camps:2010br} provides
proof of the intrinsic blackfold equations that describe fluctuations of
$r_0$ and $u^a$, and of the regularity of the black brane solutions
that result.

It remains now in \eqref{nearpbrane} to analyze the deviations of the
worldvolume metric $\gamma_{ab}$ from the flat metric $\eta_{ab}$. These
deformations can be of two kinds: intrinsic or extrinsic. Intrinsic
variations of the metric, which only involve quantities constructed out
of the worldvolume metric $\gamma_{ab}$ and its intrinsic curvature, can
appear due to background curvature or be induced by the worldvolume
embedding. However, around any given point
on the worldvolume we can choose Riemann normal coordinates so that the
intrinsic curvature shows up only
at second order in derivatives. Intrinsic metric fluctuations are
therefore negligible to the order that we work in this
paper. 

The extrinsic curvature of the worldvolume is of first order in
derivatives, and will be the main focus of this paper.

\subsection{Adapted coordinates for curved branes}
\label{subsec:fermi}

The bending of the brane in directions orthogonal to its worldvolume is
characterized by its extrinsic curvature tensor. We study the case in
which the typical extrinsic curvature radius $R$ is much larger than the
brane thickness $r_0$. At any given point on the worldvolume, the black
brane is slightly perturbed away from flatness by terms that are linear
in the fluctuations of the transverse coordinates. The effect can be
appropriately captured by employing a set of coordinates, analogous to
the Fermi normal coordinates around the trajectory of a particle, that
describe the neighbourhood of a $(p+1)$-dimensional submanifold
$\mathcal{W}_{p+1}$ of arbitrary codimension.

Fermi normal coordinates employ the idea, common to all normal
coordinates, of eliminating first derivatives of the metric around a
given point, but they are constrained to being adapted to a worldvolume
that is not geodesically embedded in the background spacetime. Thus,
these coordinates employ geodesic grids only in directions normal to
$\mathcal{W}_{p+1}$. Some of the first derivatives of the metric are not
eliminated, as they correspond to tensors that characterize the shape of
the embedding of the worldvolume.

As explained above, to first order in derivatives one can choose normal coordinates
$\sigma^a$ on $\mathcal{W}_{p+1}$ such that the connection on the
worldvolume is manifestly flat, $\Gamma_{ab}^c=0$. We denote by $y^i$
the coordinates in the orthogonal directions, in such a way that
$\mathcal{W}_{p+1}$ sits at $y^i=0$. Then
\begin{equation}
ds^2
=\eta_{ab}d\sigma^a d\sigma^b+dy_i dy^i+O(y/R)+O(\sigma^2/R_{int}^2)\,.
\end{equation}
Transverse indices $i,j=p+1,\dots, D$ are
raised, lowered and contracted using the flat metric $\delta_{ij}$.
We have introduced two characteristic lengths: $R$ for the extrinsic
curvature radius, and $R_{int}$ for the intrinsic curvature radius of the
worldvolume. Gauss-Codacci-type equations \cite{Carter:1992vb} imply that
typically $R_{int}\sim R$. In the following we will omit to specify that the
expansion
is valid up to terms $O(\sigma^2/R_{int}^2)$.

Now we extend these coordinates away from the submanifold to include
terms of first order in $y/R$. The Fermi construction adapted to the
worldvolume of a $p$-brane assigns the coordinates $(\sigma^a,y^i)$ to the
point that is reached by moving a unit affine parameter along the
geodesic with tangent $\partial/\partial y^i$ that intersects
$\mathcal{W}_{p+1}$ orthogonally at $\sigma^a$. Since $y^i$ now
parametrize geodesics, we have
\beq
\Gamma^\mu_{ij}=0\,,\quad\quad\mu=a,i\,.
\eeq
In these coordinates the extrinsic curvature tensor of the worldvolume at $y^i=0$ is
\beq
K_{ab}{}^i=\Gamma_{ab}^i=-\Gamma^c_{aj}\eta_{cb}\delta^{ji}\,,
\eeq
and is assumed to have a typical size
\beq
K_{ab}{}^i \sim R^{-1}\,.
\eeq
In the case of a particle, $p=0$, the extrinsic curvature corresponds to
its acceleration, $a^i=K_{tt}{}^i$. 

For $p$-branes with $p\geq 1$ there are additional terms
of first order in derivatives,
\beq
\omega_a{}^{i}{}_{j}=\Gamma^i_{aj}=-\Gamma^l_{ak}\delta_{lj}\delta^{ki}\,,
\eeq
which reflect the possibility of non-trivial holonomies along the
worldvolume. Including them,
the metric to first order in $y/R$ reads
\beq
ds^2=\left(\eta_{ab}-2 K_{ab}{}^i y_i\right)d\sigma^a 
d\sigma^b+2\omega_a{}^i{}_j y_i d\sigma^a dy^j+dy_i dy^i +O(y^2/R^2)\,.
\eeq

The coefficients $K_{ab}{}^i$ are
components of a tensor and cannot be removed by a coordinate transformation.
However, the transformation
\beq
y^i\rightarrow y^i-\sigma^a\omega_a{}^i{}_j y^j
+O(\sigma^2)\,,
\label{rotation}
\eeq
which is a $\sigma^a$-dependent rotation of the orthogonal
directions,
eliminates the $\omega_a{}^i{}_j$ from the metric. So, locally, we can
always set these to zero. Since our aim is to obtain local equations for
the $p$-brane, such terms never enter our analysis and we can
eliminate them. However, when one considers the global
construction of a blackfold solution, attention must be paid to possible
global obstructions. Such effects appear, for instance, in the study of
blackfolds in Taub-NUT spaces \cite{Camps:2008hb}. 

We conclude that the metric in Fermi normal coordinates for a $p$-brane
is, to first order,
\beq\label{fermi2}
ds^2=\left(\eta_{ab}-2 K_{ab}{}^i y_i\right)d\sigma^a 
d\sigma^b+dy_i dy^i +O(y^2/R^2)\,.
\eeq

This geometry characterizes completely the extrinsic metric
perturbation to first derivative order, and can be regarded as the background that surrounds the black
brane. The
slowly fluctuating metric \eqref{nearpbrane} 
under a generic extrinsic curvature perturbation takes the form
\beqa\label{extpertbrane1}
ds^2&=&\left(\eta_{ab}-2 K_{ab}{}^i y_i
+\frac{r_0^n}{r^n}u_a u_b \right)d\sigma^a 
d\sigma^b+\frac{dr^2}{1-\frac{r_0^n}{r^n}}
+r^2d\Omega^2_{(n+1)}\nonumber\\
 &&+h_{\mu\nu}(y^i)dx^\mu dx^\nu +O(r^2/R^2)\,.
\eeqa
Here $r$ is the radial coordinate orthogonal to $\mathcal{W}_{p+1}$,
\beq\label{ryi}
r=\sqrt{y_i y^i}\,,
\eeq
and $\Omega_{(n+1)}$ denotes the sphere at constant $r$ in the transverse direction, with
direction cosines $y^i/r$.
The metric corrections $h_{\mu\nu}$ depend on the parameters $r_0$ and
$u^a$, and are linear in the $K_{ab}{}^i$, hence they are of
first order in $1/R$.

We now make again use of
our restriction to locally linearized fluctuations: the perturbations induced by the
extrinsic curvature along each of the transverse directions $y^i$
decouple from each other. Thus we can deal with the deformation in each
normal direction $i$ separately, and study
the perturbations when $K_{ab}{}^i$ is non-zero along
only one direction $i=\hat i$. Introducing a direction cosine for $y^{\hat i}$ such that
\beq
y^{\hat i}=r\cos\theta\,,
\eeq
then \eqref{extpertbrane1} becomes
\beqa\label{extpertbrane}
ds^2&=&\left(\eta_{ab}-2 K_{ab}{}^{\hat i}\, r\cos\theta
+\frac{r_0^n}{r^n}u_a u_b
\right)d\sigma^a 
d\sigma^b+\frac{dr^2}{1-\frac{r_0^n}{r^n}}
+r^2d\theta^2+r^2\sin^2\theta d\Omega^2_{(n)}\nonumber\\
 &&+h_{\mu\nu}(r,\theta)dx^\mu dx^\nu +O(r^2/R^2)\,.
\eeqa
The extrinsic curvature deformations are proportional to $\cos\theta$,
and therefore are dipoles of $S^{n+1}$. Since the different multipoles
decouple from each other in a linearized analysis of perturbations, the
corrections must also be dipoles and hence of the
form\footnote{Detailed arguments for
these points have been presented in \cite{Emparan:2007wm}.}
\beq
h_{\mu\nu}(r,\theta)=\cos\theta\; \hat h_{\mu\nu}(r)\,.
\eeq
One can also show that in this case
$h_{\theta\theta}=h_{\Omega_{(n)}\Omega_{(n)}}$. In addition we can, and
shall,
choose a gauge in which $h_{r\theta}=0$.

In this way we have reduced the problem of finding the solution \eqref{extpertbrane} of the
Einstein equations to that of solving a
set of coupled ordinary differential equations for functions $\hat
h_{\mu\nu}(r)$ of the form
\beq\label{hath}
\hat h_{\mu\nu}(r)dx^\mu dx^\nu=\hat h_{ab}(r)d\sigma^a d\sigma^b+\hat h_{rr}(r)dr^2
+\hat h_{\Omega\Omega}(r)(d\theta^2+\sin^2\theta d\Omega^2_{(n)})\,.
\eeq
Some of the Einstein equations are actually
constraints, and we turn to them first.

\section{Blackfold equations}
\label{sec:exteqs}

Let us focus on the asymptotic region of \eqref{extpertbrane} at large
$r\gg r_0$, where the gravitational field that the
black brane creates is weak. If we expand \eqref{extpertbrane} to
linear order in $r_0^n$ we find 
\beqa\label{linblackbrane}
ds^2&=&\left(\eta_{ab}-2 K_{ab}{}^{\hat i}\, r\cos\theta
+\frac{r_0^n}{r^n}u_a u_b
\right)d\sigma^a d\sigma^b
+\left(1+\frac{r_0^n}{r^n}\right)dr^2
+r^2\left(d\theta^2+\sin^2\theta d\Omega^2_{(n)}\right)\nonumber\\
 &&+
\cos\theta\; \hat h_{\mu\nu}(r)dx^\mu dx^\nu+ O(r_0^{2n}/r^{2n})\,.
\eeqa
This can be written as
\beqa\label{branemetric}
ds^2&=&\lp \eta_{ab}-2 K_{ab}{}^{\hat i}\, r\cos\theta+\frac{16\pi G}{n\Omega_{(n+1)}}
\lp T_{ab}-\frac{1}{D-2}T\eta_{ab}\rp\frac{1}{r^n}\rp d\sigma^a d\sigma^b\nonumber\\
&&
+\lp 1-\frac{16\pi G}{\Omega_{(n+1)}}\frac{1}{D-2}\frac{T}{r^n}\rp
dr^2+r^2\lp d\theta^2+\sin^2\theta d\Omega^2_{(n)}\rp+\nonumber\\ 
&&+
\cos\theta\; \hat h_{\mu\nu}(r)dx^\mu dx^\nu+ O(T_{ab}^{2}/r^{2n})\,,
\eeqa
where $T_{ab}$ is the stress-energy tensor \eqref{Tab} and
$T=\eta^{ab}T_{ab}$. 
In fact, the
asymptotic form of the metric \eqref{branemetric} is generic
for any
gravitating $p$-brane, not necessarily a vacuum solution, when $T_{ab}$
is the brane stress-energy tensor (\eg the quasilocal one) measured at
$r\gg r_0$.\footnote{Leaving aside the extrinsic curvature perturbation,
the asymptotic form of this metric is more often given in isotropic
coordinates, which are obtained by changing $r\to r -16\pi G
T/(2n(D-2)\Omega_{(n+1)} r^{n-1})$.} By restricting the analysis to the
large-$r$ asymptotic region we can be generic and consider $p$-branes
with arbitrary $T_{ab}$, instead of only the case of neutral black
$p$-branes that is studied in the rest of the paper. The above form of
the metric assumes that the brane solution does not have any fields with
stress-energy components along the directions transverse to the brane
which would enter to the required order. We are effectively assuming that the
brane is sufficiently localized in its transverse directions and that there
are no external fields nor other forces acting on it. As a consequence, in this
asymptotic region we have effectively vacuum equations, even for
non-vacuum branes.

The corrections $\hat h_{\mu\nu}(r)$ in \eqref{branemetric} are
considered only to leading order in both $r/R$ and $T_{ab}/r^n$, and are
obtained by solving Einstein's equations. By direct computation of the
Einstein tensor $G_{\mu\nu}$ of
\eqref{branemetric}, one finds that the combination
\beq\label{Rrth}
G_{r\theta}-\frac{r\tan\theta}{n+1} G_{rr}=
\frac{n+2}{n+1}\frac{\sin\theta}{r^n}\frac{8\pi G}{\Omega_{(n+1)}}\, T^{ab}K_{ab}{}^{\hat i}\,
\eeq
does not involve the $\hat h_{\mu\nu}$. Thus the corresponding Einstein
equation is a constraint\footnote{This must follow from the
Gauss-Codacci equations for a surface at large constant $r$, but we have
not investigated the precise relationship in detail.} which
takes the form
\beq\label{carter}
T^{ab}K_{ab}{}^{\hat i}=0\,.
\eeq
Here ${\hat i}$ denotes an arbitrary direction transverse to the brane, so these
are the extrinsic equations of motion \eqref{exteqs}, derived for a generic
gravitating $p$-brane curved in a manner governed by the
Einstein equations. 

It may be worth noting that if these equations are not satisfied, the geometry
develops singularities on $\theta=0$ or $\theta=\pi$ at all $r$, reflecting the presence of unbalanced
stresses. These singularities are possibly conical
defects when $n=1$, like in the case of five-dimensional unbalanced black
rings, but more generally they give rise to divergent curvatures. Note, however,
that such singularities are unrelated to the possible presence of a black brane horizon.
 
In the next section we discuss the solution at all values of $r$,
without requiring that $r\gg (T_{ab})^{1/n}$.

\section{Perturbed metric solution}
\label{sec:perthor}

We return again to the particular case of a neutral black $p$-brane.
We intend to solve for the perturbed geometry, in such a way that the horizon
remains regular, for
any extrinsic perturbation that satisfies \eqref{carter} with the
stress-energy tensor \eqref{Tab}, \ie
\beq\label{carterbfeq}
n\, u^a u^b K_{ab}{}^{\hat i}=K^{\hat i}\,.
\eeq
For the remainder of this section we shall omit the transverse index
$\hat i$ from the extrinsic curvature in order to lighten the notation.
Our analysis draws heavily on the one in
\cite{Emparan:2007wm,Caldarelli:2008pz}, which we simplify and
reformulate in a worldvolume-covariant manner that applies to generic
long-wavelength extrinsic perturbations of black $p$-branes. 

The metric corrections $\hat h_{\mu\nu}(r)$ in \eqref{hath}
are
computed to linear order in the
extrinsic curvature perturbation and therefore must be proportional to worldvolume
tensor structures of order $1/R$ built out
of $K_{ab}$, $\eta_{ab}$ and
$u^a$. They can be written in terms of only five functions, in the form
\beq\label{hathab}
\hat h_{ab}(r)=K_{ab}\,\mathsf{h}_1(r)
+u^c u_{(a}K_{b)c}\,\mathsf{h}_2(r)
+ K u_a u_b\,h_\gamma(r)\,,
\eeq
\beq\label{hathr}
\hat h_{rr}(r)=K\,\lp 1-\frac{r_0^n}{r^n}\rp^{-1} h_r(r)\,,
\eeq
\beq\label{hathth}
\hat h_{\Omega\Omega}(r )=K\,r^2h_\Omega( r)\,.
\eeq

We have imposed eq.~\eqref{carterbfeq} in order to reduce the number of
independent tensor structures. In particular, the only independent
worldvolume scalar is $K$, and therefore $\hat h_{rr}$ and $\hat
h_{\Omega\Omega}$ must be proportional to it. In $\hat h_{ab}$ we could
have included a piece proportional to $K\eta_{ab}$. However, one can
easily argue that this term must vanish. Consider a brane with
directions in its worldvolume that are orthogonal to the extrinsic
curvature. Terms $\propto \eta_{ab}$ deform the brane in these
directions, but there cannot be any such deformations since there is no
force to induce them --- the dynamics along these directions can be
consistently truncated out. It might seem that, by a
similar argument, the function $h_\gamma(r)$ that multiplies $K u_a u_b$
should vanish, too. However, this term can be generated by a residual
diffeomorphism. This is the freedom to change coordinates to first order
in $K$ as
\beqa\label{rthgauge}
r\to r+K\cos\theta\,\gamma(r) \,,\qquad
\theta\to\theta+K\sin\theta \,
\int^r dr' \frac{\gamma(r')}{{r'}^2\lp 1-\frac{r_0^n}{{r'}^n}\rp}\,,
\eeqa
under which $\mathsf{h}_1$ and $\mathsf{h}_2$
remain invariant but
\beqa
h_\gamma&\to&h_\gamma -n\frac{r_0^n}{r^{n+1}} \gamma(r)\,,\nonumber\\
h_r&\to& h_r+2\gamma'(r)-n\frac{r_0^n}{r^{n+1}}
\frac{\gamma(r)}{1-\frac{r_0^n}{r^n}}\,,\\
h_\Omega'&\to& h_\Omega'+2\frac{\gamma'(r)}{r}+2\frac{r_0^n}{r^{n+2}}
\frac{\gamma(r)}{1-\frac{r_0^n}{r^n}}\nonumber\,.
\eeqa
In the way we have written $\hat h_{rr}(r)$, in order that the
horizon remains at $r=r_0$ these transformations are constrained to satisfy
$\gamma(r_0)=0$. 

We might use this freedom for, \eg\ setting
$h_\gamma$ to zero. However, since it will turn out that horizon regularity demands
that $h_\gamma(r_0)\neq 0$, it is preferrable to leave it as
a gauge-dependent
variable and introduce two functions 
\beqa\label{hrandhtheta}
\mathsf{h}_{r}&=&h_r+\frac{2}{n}r_0\left(\frac{r^{n+1}}{r_0^{n+1}}h_\gamma\right)^\prime-\frac{h_\gamma}{1-\frac{r_0^n}{r^n}}\,,\nonumber\\
\mathsf{h}'_{\Omega}&=&h_\Omega'+\frac{2}{n}\frac{r_0}{r}\left(\frac{r^{n+1}}{r_0^{n+1}}h_\gamma\right)^\prime+\frac{2}{n\, r}\frac{h_\gamma }{1-\frac{r_0^n}{r^n}}
\eeqa
that are invariant under \eqref{rthgauge}. 

The
general extrinsic perturbations are then characterized by the
four gauge-invariant functions $\mathsf{h}_1$, $\mathsf{h}_2$,
$\mathsf{h}_r$ and $\mathsf{h}_\Omega$, subject to suitable boundary conditions.
In order to maintain the asymptotic behavior in \eqref{linblackbrane} we
require that $\hat h_{\mu\nu}(r)=O(r^{-n+1})$ as $r\to\infty$. This constrains the
residual gauge symmetry \eqref{rthgauge} to transformations with
$\gamma'=O(r^{-n+1})$. For the gauge-invariant functions, the asymptotic
behavior is then fixed to be
\beq\label{hasymp}
\mathsf{h}_{1,2}=O(r^{-n+1})\,,\qquad \mathsf{h}_{r,\Omega}=r^2 O(r^{-1})\,.
\eeq

In order to find the solution it suffices to solve the equations for some particular
configuration of the extrinsic curvature and velocity that excites the
three structures in \eqref{hathab}. 
This has been done already in
\cite{Emparan:2007wm,Caldarelli:2008pz}, and in
appendix~\ref{app:soln} we give some details on how to convert the
results to our set up.

Some of the solutions to the linear perturbation equations with  the
asymptotics \eqref{hasymp}, while
satisfying the extrinsic blackfold equations \eqref{carterbfeq}, give
singular behavior of $\mathsf{h}_{1,2}$ on the horizon and thus we discard them.
The solution where $\mathsf{h}_{1,2}$ remain finite at $r=r_0$
and which satisfies \eqref{hasymp}
is unique and reads
\begin{subequations}\label{hsoln}
\beq
\mathsf{h}_1=2r-A\; P_{1/n}\left(2\frac{r^n}{r_0^n}-1\right)\,,
\eeq
\beq
\mathsf{h}_2=
-A\; 
\frac{r_0^{n}}{r^n}\left[P_{1/n}\left(2\frac{r^n}{r_0^n}-1\right)
+P_{-1/n}\left(2\frac{r^n}{r_0^n}-1\right)\right]\,,
\eeq
\beq
\mathsf{h}_r=
\frac{n+1}{n^2\left(1-\frac{r_0^n}{r^n}\right)}
\left[\lp \frac{n}{n+1}-2\frac{r_0^n}{r^n}\rp (2r-\mathsf{h}_1) -\mathsf{h}_2
\right]\,,
\eeq
\beq
\mathsf{h}^\prime_\Omega=\frac{1}{nr\left(1-\frac{r_0^n}{r^n}\right)}
\left( 2r-\mathsf{h}_1 +\frac{n+2}{2n} \mathsf{h}_2\right)\,,
\eeq
\end{subequations}
where $P_{\pm 1/n}(x)$ are Legendre functions 
and we have defined a constant
\beq\label{Adef}
A=2 r_0\frac{\Gamma\left(\frac{n+1}{n}\right)^2}{\Gamma\left(\frac{n+2}{n}\right)}\,.
\eeq

Having found these functions, the complete metric is specified by making
a choice of the gauge-dependent function $h_\gamma(r)$. This must be subject to
the asymptotic boundary condition\footnote{In general this modifies the 
asymptotic Fermi normal frame by introducing terms in
$h_{rr}$, $h_{\Omega\Omega}$ $\sim K r\cos\theta$. If we
we want to avoid them we must require $h_\gamma=\frac{1}{2}\frac{r_0^n}{r^{n-1}}+O(r^{-1-n})$.}
\beq\label{hginf}
h_\gamma=O(r^{-n+1})\,.
\eeq
In the next section we shall see that the value of $h_\gamma$ at 
$r=r_0$ is constrained
by the requirement of horizon regularity.
Indeed, it must be fixed in order to cancel in $h_r$ and $h'_\Omega$ the
singular behavior that $\mathsf{h}_r$ and $\mathsf{h}^\prime_\Omega$ have at that point.

\section{Horizon regularity}
\label{sec:horreg}

Now we exhibit Eddington-Finkelstein coordinates for the
perturbed solution which make the horizon manifestly regular.

It is convenient to organize the worldvolume tensor structures into components
parallel and orthogonal to the velocity. Using the orthogonal projector
\beq\label{Pab}
P_{ab}=\eta_{ab}+u_a u_b\,,
\eeq 
and always assuming that the constraint \eqref{carterbfeq} is satisfied,
the perturbed metric takes the form
\beqa\label{corrmet2}
ds^2&=&\left[ 
\eta_{ab}+\frac{r_0^n}{r^n}u_a u_b +\lp
K_{cd} P^c{}_a P^d{}_b\,\mathsf{g}_1 
+ u^c K_{cd}P^d{}_{a}u_{b}\,\mathsf{g}_2
+K u_a u_b\, g_\gamma
\rp\cos\theta
\right]d\sigma^a d\sigma^b\nonumber\\
&&+\lp 1+K h_r\cos\theta\rp\frac{dr^2}{1-\frac{r_0^n}{r^n}} +r^2\lp 1
+K h_\Omega\cos\theta\rp (d\theta^2+\sin^2\theta d\Omega^2_{(n)})\,,
\eeqa
where the new functions are obtained from the ones in the previous
section as
\beqa
\mathsf{g}_1&=&\mathsf{h}_1-2r\,,\\
\mathsf{g}_2&=&
4r -2\mathsf{h}_1+\mathsf{h}_2\,,\\
g_\gamma&=&h_\gamma+\frac{1}{n}(\mathsf{h}_1-\mathsf{h}_2-2r)\,.
\eeqa
The $\mathsf{g}_{1,2}$ are invariant under
the gauge transformation \eqref{rthgauge}, while $g_\gamma$ is changed by it.
The extrinsic curvature terms of the asymptotic Fermi frame have been
absorbed in these functions.

Expanding the solutions \eqref{hsoln} around $r=r_0$, we find that
\beqa
\mathsf{g}_1&=&-A+O(r-r_0)\,,\\
\mathsf{g}_2&=&\frac{2A(n+1)}{r_0}(r-r_0)+O(r-r_0)^2
\,,\\
g_\gamma&=&h_\gamma(r_0)+\frac{A}{n}+O(r-r_0)\,,\\
h_r&=&\lp h_\gamma(r_0)+\frac{A}{n}\rp\frac{r_0}{n(r-r_0)}+O(r-r_0)^0\,,\\
h_\Omega'&=&-\lp h_\gamma(r_0)+\frac{A}{n}\rp\frac{2}{n^2(r-r_0)}+O(r-r_0)^0\,.
\eeqa
From the last two expressions it is apparent that in order to preserve
regularity at $r=r_0$ it is necessary to set 
\beq\label{hghor}
h_\gamma(r_0)=-\frac{A}{n}\,,
\eeq
which also makes $g_\gamma$ vanish at $r_0$. This
implies that $u^a u^b g_{ab}$ is zero there. 

Condition \eqref{hghor} turns out to be sufficient for regularity.
When it is imposed, one has
\beqa
g_\gamma&=&\lp\frac{A}{r_0}\frac{1-n-2n^2}{n^2}+h'_\gamma(r_0)\rp (r-r_0)+O(r-r_0)^2\,,\\
h_r&=&-\frac{r_0}{n}g'_\gamma(r_0)+O(r-r_0)\,.
\eeqa
Also, $h_\Omega(r)=\int dr\, h_\Omega'(r)$ is regular at $r=r_0$ for any value of the integration constant.

Performing a change to Eddington-Finkelstein coordinates in the form
\beq
dv=-u_a d\sigma^a+\frac{dr}{1-\frac{r_0^n}{r^n}}\,,
\eeq
the metric becomes
\beqa
ds^2&=&2\lp 1+K h_r(r_0)\cos\theta\rp dr dv +
2A\frac{n+1}{n} u^c K_{cd}P^d{}_a \cos\theta\, d\sigma^a\, dr
\nonumber\\
&& 
+\lp P_{ab}-A\, K_{cd} P^c{}_a P^d{}_b\cos\theta\rp d\sigma^a d\sigma^b\\
&&+r^2\lp 1+K h_\Omega(r_0)\cos\theta\rp (d\theta^2+\sin^2\theta d\Omega^2_{(n)})
+O(r-r_0)\,,\nonumber
\eeqa
which is manifestly non-singular at $r=r_0$. 

Since the perturbation is purely dipolar, neither the horizon
temperature nor its entropy-density receive any corrections
\cite{Emparan:2007wm}. The horizon velocity is also uncorrected, since
the generators of the horizon are the orbits of $u^a$. The values of
$h_r(r_0)$ and $h_\Omega(r_0)$ are gauge-dependent and unconstrained by
regularity requirements. Both of them could be set to zero if desired.
The only physical effect on the geometry of the horizon is the distortion of the
worldvolume by the extrinsic curvature in directions orthogonal to $u^a$.

\section{Complete corrected black brane geometry}
\label{sec:completemetric}

For reference and further use, we compile the results
for all the perturbations, intrinsic and extrinsic, in a
manifestly covariant form in the parallel and transverse directions.

The metric for the fluctuating, curved black $p$-brane takes the form of
\eqref{nearpbrane},
with worldvolume metric
\beq\label{gammaab}
\gamma_{ab}=\eta_{ab}-2 K_{ab}{}^i(\sigma)\, r\cos\theta_i+O(r^2/R^2)\,.
\eeq
Here $\cos\theta_i=y_i/r$ are
direction cosines in $S^{n+1}$ which parametrize the transverse
directions and satisfy
\beq
\cos\theta_i\cos\theta^i=1\,,\qquad d(\cos\theta_i)d(\cos\theta^i)=d\Omega_{(n+1)}^2\,.
\eeq
The $n+2$ indices $i=p+1,\dots, D$ 
are summed over, when repeated, in all these equations.

The $K_{ab}{}^i(\sigma)$, $r_0(\sigma)$ and $u^a(\sigma)$ must solve the
blackfold equations
\beq\label{binteqs}
\dot u_a+\frac{1}{n+1}\vartheta u_a=\partial_a \ln r_0\,,
\eeq
\beq
n K_{ab}{}^i u^a u^b= K^i\,,
\eeq
where
\beqa
\dot u_{a}=u^b D_b u_a\,,\qquad
\vartheta =D_a u^a\,
\eeqa
are the acceleration and expansion of the effective fluid's velocity.
Then, every solution to these equations determines completely (up to the
gauge freedom discussed in the previous section) a set of
`bulk' corrections $h_{\mu\nu}$. We split these into 
fluid corrections (functions $f_{\mu\nu}$) and
extrinsic corrections (functions $h_{\mu\nu}{}^i$), such that the complete metric is
\beqa
ds^2&=&\left(\eta_{ab}-2 K_{ab}{}^i(\sigma) \, r\cos\theta_i
+\frac{r_0(\sigma)^n}{r^n}u_a(\sigma) u_b(\sigma) \right)d\sigma^a 
d\sigma^b+\frac{dr^2}{1-\frac{r_0(\sigma)^n}{r^n}}
+r^2d\Omega^2_{(n+1)}\nonumber\\
&&+
\lp f_{ab}(r)+h_{ab}{}^i(r)\cos\theta_i\rp d\sigma^a d\sigma^b
+2 f_{ar}(r)d\sigma^a dr
\nonumber\\
&&+\lp f_{rr}(r)+h_{rr}{}^i(r)\cos\theta_i \rp dr^2
+h_{\Omega\Omega}{}^i(r)\cos\theta_i\, d\Omega_{(n+1)}^2\,.
\eeqa
The fluid fluctuations do not deform the $S^{n+1}$, hence
$f_{\Omega\Omega}$ are zero.

\paragraph{Fluid corrections.} These have been derived in
\cite{Camps:2010br}. We write them as
\beqa
f_{ab}(r) &=& 
\vartheta u_a u_b\,\mathsf{f}_1(r)
+\left(\sigma_{ab}+\frac1p \vartheta P_{ab}\right)\,\mathsf{f}_2(r )
\label{fab}\,,\\
f_{ar}(r)&=&\vartheta u_a\,\mathsf{f}_3(r)+\dot u_a \,\mathsf{f}_4(r)\,,\\
f_{rr}(r) &=& \vartheta\,\lp 1-\frac{r_0^n}{r^n}\rp^{-1} \mathsf{f}_r(r)\,.
\eeqa
The derivatives of $r_0$ have been eliminated through the equations
\eqref{binteqs}. Here $P_{ab}$ is the orthogonal projector \eqref{Pab}, $\sigma_{ab}$ is the shear of the velocity flow,
\beq\label{sab}
\sigma_{ab}+\frac1p \vartheta P_{ab}=P_a{}^c P_b{}^dD_{(c}u_{d)} \,,
\eeq
and the radial functions are
\beqa
\mathsf{f}_1( r)&=& \frac{r_0}{n(n+1)} \left(2-(n+2)\frac{r_0^n}{r^n}\right)\ln\left(1-\frac{r_0^n}{r^n}\right)\,,\\
\mathsf{f}_2( r)&=& \frac{2\,r_0}{n}\ln\left(1-\frac{r_0^n}{r^n}\right)\,,\\
\mathsf{f}_3( r)&=& \frac{r_0}{n+1}\frac{1}{1-\frac{r_0^n}{r^n}}\left[\left(\frac{n+1}{n}\frac{r_0^n}{r^n}-\frac1n\right)
\ln\left(1-\frac{r_0^n}{r^n}\right)-\frac{r_0^n}{r^n}\left(n\frac{r_*}{r_0}+1\right)\right]+\delta_{n,1}\,,\quad\\
\mathsf{f}_4( r)&=& \frac{r_*-r}{1-\frac{r_0^n}{r^n}}-\delta_{n,1}r_0\ln\frac{r_0}{r}\,,\\
\mathsf{f}_r( r)&=& \frac{r_0}{n+1}\frac{1}{1-\frac{r_0^n}{r^n}}\frac{r_0^n}{r^n}\left(2-\ln\left(1-\frac{r_0^n}{r^n}\right)\right) \,,
\eeqa
where
\beq
r_*=\int\frac{dr}{1-\frac{r_0^n}{r^n}}={}_2F_1\left(1,-\frac{1}{n};\frac{n-1}{n};\frac{r_0^n}{r^n}\right)\,r\,.
\eeq

Observe that the structure $u_{(a}\dot u_{b)}$ might have appeared in $f_{ab}$,
but it turns out not to contribute.

\paragraph{Extrinsic corrections.} These are the ones obtained in
section~\ref{sec:perthor}. Now we simply include the index $i$ for all
possible directions orthogonal to the worldvolume,
\beqa
h_{ab}{}^i(r)&=&K_{ab}{}^i\,\mathsf{h}_1(r)
+u^c u_{(a}K_{b)c}{}^i\,\mathsf{h}_2(r)
+ K{}^i u_a u_b\,h_\gamma(r)\,,\\
h_{rr}{}^i(r)&=&K{}^i\,\lp 1-\frac{r_0^n}{r^n}\rp^{-1} h_r(r)\,,\\
h_{\Omega\Omega}{}^i(r)&=&K{}^i\,r^2h_\Omega(r)\,,
\eeqa
with the functions $\mathsf{h}_{1,2}$, $h_r$, $h_\Omega$ as
given in \eqref{hrandhtheta}, \eqref{hsoln}. The function $h_\gamma$ is
only constrained to satisfy \eqref{hginf} and
\eqref{hghor}, and the integration constant in $h_\Omega$ can be chosen
arbitrarily. 

Eq.~\eqref{corrmet2} gives an alternative form for the
metric, with a decomposition of $g_{ab}$ based on directions parallel and
orthogonal to $u^a$.

\section{Discussion}
\label{sec:discuss}

We have shown that the effective blackfold formalism of
\cite{Emparan:2009at} can be fully derived from the Einstein equations
for a deformed black brane, and results in geometries that are regular
on and outside the black brane horizon. In particular this proves that
all the new solutions in \cite{Emparan:2009vd} are, to leading
derivative order, \textit{bona fide} black holes. In fact our solution
applies to neutral black branes in any background with typical length scales
much larger than $r_0$. This includes (Anti-)deSitter
backgrounds with cosmological radius $\sqrt{|\Lambda|}\gg r_0$ and black
hole backgrounds with horizon sizes $R_h\gg r_0$
\cite{Caldarelli:2008pz,Armas:2010hz}.

The extension to higher derivative orders is presumably technically
complicated, since the different types of perturbations couple at the
second and higher orders. Particular subsets of perturbations may be
more amenable to study, but one should bear in mind that all the
problems that one encounters in the analysis of self-force for small
black holes \cite{Poisson:2011nh} are probably exacerbated for extended
objects like black $p$-branes.

Our study motivates the discussion of several issues:

\paragraph{The metric at all scales.} Having obtained the solution in
section~\ref{sec:completemetric} for the deformed black $p$-brane in the
region where $r\ll R$, finding the metric valid at all radii only
requires to complete a straightforward matched asymptotic expansion.

The construction is as follows. The black brane modifies the geometry of
the background spacetime that it lives in. To first order for a thin
brane, the correction is the linearized field sourced by a stress-energy
tensor $T_{ab}$ localized on the worldvolume at $x^\mu=X^\mu(\sigma^a)$.
Linearized gravity is a well studied subject and, in \eg\ Minkowski or
AdS spacetimes the required solution can be readily expressed in terms
of integrals of known Green's functions over the
given source.

The solution to this problem in linearized gravity gives the `far zone'
geometry, in the region where the brane thickness is negligible, $r\gg
r_0$. At small $r$, we know that this must match the geometry
\eqref{linblackbrane} with specific values for $K_{ab}{}^i$, since this
form is valid for $r_0\ll r\ll R$ (the `overlap zone'). The solution in
section~\ref{sec:completemetric} extends this geometry into the entire
region $r\ll R$ --- the `near zone' --- for any $K_{ab}{}^i$. So, once
we make explicit the `far zone' metric for the specific brane source,
the results in this paper give the complete metric for the bent
black brane.

\paragraph{Viscous fluid, elastic solid.} The solution for the strained
black brane in section \ref{sec:completemetric} can be used
to compute the stresses induced on the black brane as derivative corrections to
$T_{ab}$. From these stress-strain relations we can extract
linear-response coefficients, namely, the effective fluid viscosities
and moduli of elasticity. These have been calculated in
\cite{Camps:2010br} and \cite{Armas:2011uf} respectively. We present
the resulting stress-energy tensor for neutral black
$p$-branes in the form
\beq\label{finalstresstensor}
T_{ab}=\Big(\varepsilon u_a u_b +P P_{ab}-2\eta\sigma_{ab}-\zeta\vartheta P_{ab}+\tilde{Y}_{ab}{}^{cd}K_{cd}{}^i\partial_i\Big)
\delta^{(n+2)}_\perp\left(x^i-X^i(\sigma^a)\right)\,,
\eeq
where $\delta_\perp^{(n+2)}$ localizes in directions transverse to the
worldvolume, and the constitutive relations are
\beq
\frac{\varepsilon}{n+1}=-P=\frac{\Omega_{(n+1)}r_0^{n}}{16\pi G}\,,
\eeq
\beq
\eta=\frac{\Omega_{(n+1)}r_0^{n+1}}{16\pi G}\,,
\qquad\zeta=2\eta\left(\frac{1}{p}+\frac{1}{n+1}\right)\,,
\eeq
and
\beqa
\tilde{Y}_{ab}{}^{cd}&=&\frac{\Omega_{(n+1)}r_0^{n}}{16\pi G}\frac{n\tan(\pi/n)}{4\pi}A^2
\left(\frac{1}{n+2}\delta_{(a}{}^c\delta_{b)}{}^d+2u_{(a}\delta_{b)}{}^{(c}u^{d)}+\frac{3n+4}{n+2} u_a u_b u^c u^d
\right)\nonumber\\
&&+\,\tilde{k}\,\frac{n\tan(\pi/n)}{4\pi}A^2\left[\left(\varepsilon u_a u_b+P P_{ab}\right)\eta^{cd}+\eta_{ab}\left(\varepsilon u^c u^d+P P^{cd}\right)\right]\,.
\eeqa
The expression for the elasticity moduli $\tilde{Y}_{ab}{}^{cd}$ is symmetric under $\{
ab\}\leftrightarrow\{cd\}$ and differs from \cite{Armas:2011uf} by terms that vanish in $T_{ab}$ when the equations of motion are imposed.
The free constant $\tilde{k}$ comes from the residual diffeomorphism \eqref{rthgauge} and parametrizes an ambiguity in the
specification of the worldvolume surface.\footnote{
The dimensionally correct Young modulus $Y_{ab}{}^{cd}$ as defined in
\cite{Armas:2011uf} is
\beq
Y_{ab}{}^{cd}=\frac{(n+2)(n+4)}{r_0^{n+4}}\frac{\Omega_{(n)}}{\Omega_{(n+1)}}\tilde{Y}_{ab}{}^{cd}\,,\nonumber
\eeq
and is valid for $n>2$.
The gauge $\tilde{k}_2=0$ for the dipole in  \cite{Armas:2011uf} corresponds to
$\tilde{k}=-\frac{(n+1)(n+4)}{n^2(n+2)}$ here.}

It is worth stressing that these values of the linear-response
coefficients are fixed by the requirement of regularity of the horizon
in the perturbed geometry, as imposed in
sec.~\ref{sec:perthor} and in \cite{Camps:2010br}.

\paragraph{Blackfold boundaries.} Our analysis applies to
black branes whose thickness $r_0$ remains finite. However, at blackfold
boundaries the thickness vanishes \cite{Emparan:2009at}. The current
evidence is that when this happens as a consequence of the velocity $u^a$
becoming lightlike, one obtains a regular horizon
\cite{Emparan:2009cs,Emparan:2009vd,Caldarelli:2010xz}. If, instead, it
is due to the blackfold intersecting a horizon of the background, then a
conifold-type singularity appears \cite{Emparan:2011ve}. These phenomena
remain the only aspect of the effective blackfold theory for which we
still do not have a general understanding.

\paragraph{Other black branes and AdS/CFT.} 
The effective theory of
blackfolds in \cite{Emparan:2009at}
has been extended to charged black branes
\cite{Grignani:2010xm,Caldarelli:2010xz,Emparan:2011hg}, and for these,
the analysis of section~\ref{sec:perthor} (but not that of
section~\ref{sec:exteqs}) needs to be redone since it involves a different
set of field equations. We do not expect any difficulties of principle
in doing it.

D-branes are a particular case of the charged branes analyzed with
blackfold methods in \cite{Emparan:2011hg}, and one may study their
bending by an extension of our approach. Some of these D-branes admit a
decoupling limit in which the flat, extremal brane geometries become of
type AdS$_{p+2}\times S^{n+1}$, famously dual to conformal field
theories. Extrinsic perturbations such as we have studied give rise to
deformations of the $S^{n+1}$, and correspond to giving certain
vev's for the scalar fields in the R-symmetry group (locally $SO(n+2)$)
of the dual conformal theory. It may be of interest to compute this effect.

\section*{Acknowledgments}

JC thanks the University of Barcelona for warm hospitality. We are also
grateful to the String Theory Group at University of Amsterdam for
hospitality in the last stages of this work, during the workshop on
``Holographic Fluids". JC was supported by the STFC Consolidated Grant ST/J000426/1. RE was
supported by MEC FPA2010-20807-C02-02, AGAUR 2009-SGR-168 and CPAN
CSD2007-00042 Consolider-Ingenio 2010. 

\appendix

\section{Solution of Einstein's equations}
\label{app:soln}

The results we presented in sec.~\ref{sec:perthor} are the worldvolume
covariantization of the solutions in \cite{Emparan:2007wm,Caldarelli:2008pz}.
In this appendix we give some details on how the
functions $\mathsf{h}$ of sec.~\ref{sec:perthor} are related
to those in  \cite{Emparan:2007wm,Caldarelli:2008pz}.

Refs.~\cite{Emparan:2007wm,Caldarelli:2008pz} study the case $p=1$, with a
boost parametrized by rapidity $\alpha$ such that $u_a=(-c_\alpha,-s_\alpha)$,
and with extrinsic curvature $K_{ab}=\textrm{diag}(C_t,-C_z)$. In this
case the
other tensor structure in eqs.~\eqref{hathab} is
\beq
u^c u_{(a}K_{b)c}=
-\left(\begin{array}{cc}	
C_t\, c_\alpha^2 & \frac{C_t +C_z}{2}c_\alpha s_\alpha\\
 \frac{C_t +C_z}{2}c_\alpha s_\alpha& C_z\, s_\alpha^2
\end{array}\right)\,.
\eeq

The functions $\mathsf{A}$ and $\mathsf{B}$ in
\cite{Emparan:2007wm,Caldarelli:2008pz} are related to our
$\hat{h}_{ab}$ as
\beq\label{A(h)}\begin{split}
\mathsf{A}=&\hat{h}_{tt}-\frac{c_\alpha^2}{s_\alpha^2}\hat{h}_{zz}
\\=&\left(K_{tt}-\frac{c_\alpha^2}{s_\alpha^2}K_{zz}\right) 
(\mathsf{h}_1-2r)+\left(u^c u_t K_{tc}-\frac{c_\alpha^2}{s_\alpha^2}u^c u_z K_{zc}\right)\mathsf{h}_2\\
=&\left(C_t+\frac{c_\alpha^2}{s_\alpha^2}C_z\right)(\mathsf{h}_1-2r)+c_\alpha^2\left(-C_t+C_z\right)\mathsf{h}_2\,,\\
\end{split}\eeq
\beq\label{B(h)}\begin{split}
\mathsf{B}=&\frac{1}{c_\alpha s_\alpha}\hat{h}_{tz}-\frac{1}{s_\alpha^2}\hat{h}_{zz}\\=&
\left(\frac{1}{c_\alpha s_\alpha}K_{zz}K_{tz}-\frac{1}{s_\alpha^2}K_{zz}\right) 
(\mathsf{h}_1-2r)+\left(\frac{1}{c_\alpha s_\alpha}u^c u_{(t} K_{z)c}-\frac{1}{s_\alpha^2}u^c u_z K_{zc}\right)\mathsf{h}_2\\
=&\frac{C_{z}}{s_\alpha^2} (\mathsf{h}_1-2r)
+\frac{C_z-C_t}{2}\mathsf{h}_2\,.
\end{split}\eeq
The function $h_\gamma$ does not appear here, which is equivalent to
absence of the gauge-dependent function $c(r)$ in
\cite{Emparan:2007wm,Caldarelli:2008pz}.

Refs.~\cite{Emparan:2007wm,Caldarelli:2008pz} solved the linearized
equations for a generic dipole perturbation of a black string and obtained
four independent solutions. Two of them
diverge at $r=r_0$ and are discarded, and the ones that remain finite
at $r=r_0$ are
\beq\begin{split}
u_1&= {}_2 F_1\left(-\frac{1}{n},-\frac{n+1}{n};1;1-\frac{r_0^n}{r^n}\right)r\,,\\
u_2&= {}_2 F_1\left(-\frac{1}{n},\frac{n-1}{n};1;1-\frac{r_0^n}{r^n}\right)\frac{r_0^n}{r^{n-1}}\,.
\end{split}\eeq
In terms of these, the solution for $\mathsf{h}_{1}$ and
$\mathsf{h}_{2}$ that satisfies the asymptotic boundary
conditions \eqref{hasymp} is
\beq
\mathsf{h}_1=-\frac{A}{r_0}\frac{n+1}{n+2}\left(u_1+\frac{1}{n+1}u_2\right)+2r\,,\quad\quad
\mathsf{h}_2=-\frac{2A}{r_0}u_2\,,
\label{h(u)}\eeq
where $A$, defined in \eqref{Adef}, gives a simpler form for the
coefficient of $u_1$ than the constant $A_1$ in
\cite{Emparan:2007wm,Caldarelli:2008pz}. 
It is straightforward to check that substituting \eqref{h(u)} in
\eqref{A(h)} we recover the correct expressions for $\mathsf{A}$ and
$\mathsf{B}$. We
omit the details of how the functions $\mathsf{h}_{r}$ and
$\mathsf{h}_{\Omega}'$ are similarly obtained from $\mathsf{F}$ and
$\mathsf{G}'$ in \cite{Emparan:2007wm,Caldarelli:2008pz}. 

The hypergeometric functions in the solution are actually
Legendre functions $P_{\nu}(x)$ with index $\nu=\pm 1/n$ and argument
$x=2(r/r_0)^n- 1$. A simple example of such relations is
\beq
{}_2 F_1\left(-\frac{1}{n},-\frac{1}{n};1;1-
\frac{r_0^n}{r^n}\right)=\frac{r_0}{r}P_{1/n}\left(2\frac{r^n}{r_0^n}-
1\right)\,.
\eeq
This and other similar expressions allow to rewrite the $\mathsf{h}$ functions
as in \eqref{hsoln}.
The Legendre functions $P_{\nu}(x)$ become Legendre polynomials when
$\nu\in\mathbb{N}$, which in the present instance can only be when $n=1$.

\end{document}